\documentclass[notitlepage,nofootinbib,preprintnumbers,amssymb,superscriptaddress,eqsecnum]{revtex4-1}
\usepackage{amsfonts,amssymb,amsmath,graphicx,color,bm}
\definecolor{ultramarine}{rgb}{0.07, 0.04, 0.56}
\definecolor{cadmiumgreen}{rgb}{0.0, 0.42, 0.24}
\definecolor{indigo(dye)}{rgb}{0.0, 0.25, 0.42}
\usepackage[linktocpage=true]{hyperref}
\hypersetup{
colorlinks=true,
citecolor=ultramarine,
linkcolor=cadmiumgreen,
urlcolor=indigo(dye),
}

\newcommand{\f}[2]{\frac{#1}{#2}}  
\newcommand{\mk}[1]{\left( #1 \right)}  
\newcommand{\kk}[1]{\left[ #1 \right]}  
  
\newcommand{\be}{\begin{equation}}  
\newcommand{\ee}{\end{equation}}
\newcommand{\bem}{\begin{pmatrix}}
\newcommand{\eem}{\end{pmatrix}}
\newcommand{\Mpl}{M_{\rm Pl}}
\newcommand{\pa}{\partial}
\newcommand{\hb}{\hat\beta_0}
\newcommand{\hbc}{\hat\beta_{0, {\rm c}}}

\begin{document}

\preprint{YITP-18-105, IPMU18-0175}

\title{  
Shape dependence of spontaneous scalarization 
}

\author{Hayato Motohashi}
\affiliation{Center for Gravitational Physics, Yukawa Institute for Theoretical Physics, Kyoto University, Kyoto 606-8502, Japan}

\author{Shinji Mukohyama}
\affiliation{Center for Gravitational Physics, Yukawa Institute for Theoretical Physics, Kyoto University, Kyoto 606-8502, Japan}
\affiliation{Kavli Institute for the Physics and Mathematics of the Universe (WPI), The University of Tokyo, 277-8583, Chiba, Japan}
\affiliation{Institut Denis Poisson, UMR - CNRS 7013, Universit\'{e} de Tours, Parc de Grandmont, 37200 Tours, France}

\begin{abstract}
 Spontaneous scalarization is an interesting mechanism for modification of gravity by nonminimal coupling of a scalar field to matter or curvature invariants in the context of scalar-tensor theories, and its onset is signaled by linear instability of the scalar field around the corresponding general relativity solution. We thus perform the linear stability analysis of the scalar field about general relativity solutions and highlight a crucial difference between a spherically symmetric profile and a planar symmetric profile. We clarify that the critical value for the instability is sensitive to the morphology and that the spontaneous scalarization occurs much more easily with the planar symmetric shape than with the spherically symmetric shape. 
\end{abstract}

\maketitle  


\section{Introduction}
\label{sec:intro}

Spontaneous scalarization has been attracting much attention recently as a mechanism for local modification of gravity.  This phenomenon was found in \cite{Damour:1993hw} for the canonical scalar-tensor theory without potential but with coupling to the matter Lagrangian via a conformally transformed metric $\Omega^2(\phi)g_{\mu\nu}$. Later, the model was generalized for a massive scalar field~\cite{Chen:2015zmx,Ramazanoglu:2016kul}. In these models, the evolution of the scalar field is affected by the stress energy tensor of the matter component due to the nonminimal coupling of the scalar field. Therefore, by designing the dependence of the conformal factor on the scalar field, one can modify gravity in high density regions. Specifically, with $\Omega_{,\phi}=0$ and $\Omega_{,\phi\phi}<0$ at $\phi=\phi_0$, the scalar field in a high density region exhibits tachyonic instability causing the spontaneous scalarization, which leads to interesting phenomenology~\cite{Berti:2015itd}. The generalization to those theories in which the scalar field couples to matter via a disformally transformed metric was also investigated~\cite{Minamitsuji:2016hkk}. Furthermore, the spontaneous scalarization in the Einstein-scalar-Gauss-Bonnet theory was also investigated~\cite{Doneva:2017bvd,Silva:2017uqg,Antoniou:2017acq},  in which the scalarization is triggered by the tachyonic mass due to the coupling between a scalar field and the Gauss-Bonnet term. 
It is also shown that the Schwarzschild-Newman-Tamburino-Unti solution can get scalarized in models 
with a nonminimal couplig to either the Gauss-Bonnet or the Chern-Simons terms~\cite{Brihaye:2018bgc}.
The spontaneous scalarization of the electrically charged 
black holes in the presence of nonminimal couplings between a scalar field and the Maxwell invariant is also explored~\cite{Herdeiro:2018wub}.

The study of spontaneous scalarization in spherically symmetric configurations is somewhat mature. To be a realistic model, however, it is important to investigate it with other shapes of the matter distribution. The main purpose of the present paper is to reveal the shape dependence of the spontaneous scalarization in a concise manner. To highlight the shape dependence, we compare a spherically symmetric spacetime and a planar symmetric spacetime. The occurrence of the spontaneous scalarization is manifested as linear instability of perturbation of the scalar field about the corresponding general relativity (GR) solution with $\phi=\phi_0$~\cite{Harada:1997mr}. In this framework the linear perturbation of the scalar field does not affect the background GR solution since the stress energy tensor of the scalar field and the $\phi$-dependence of the conformal factor start at the second order of $\delta\phi \equiv \phi - \phi_0$. Hence the evolution of the scalar perturbation $\delta\phi$ at the linear order is governed by the perturbed Klein-Gordon equation with the fixed GR background metric, which takes the form of a Schr\"odinger equation with an effective potential and energy after mode decomposition. As we shall see below, the instability of the scalar perturbation, i.e.\  the spontaneous scalarization, corresponds to the existence of a bound state with negative energy in the Schr\"odinger problem.

While in general the form of the effective potential is not simple, it is expected that we can acquire some insight from a simpler Schr\"odinger problem with the square well potential~\cite{1968quantum}. Let us therefore consider the Schr\"odinger equation 
\be \label{Sch-eq}
\mk{-\f{d^2}{dx^2} + U } \psi = E \psi\,,
\ee
with the square well potential
\be
 U(x) =
 \begin{cases}
  -U_0 & (|x|<D)\,, \\
  0 & (|x|>D)\,.
 \end{cases}
\ee
Here, $x$ is the radial coordinate. Since the potential is an even function, without loss of generality we can assume that $\psi(x)$ is either odd or even. The planar symmetric case amounts to $-\infty < x < \infty$ whereas the spherically symmetric case amounts to $0 < x < \infty$. It is well known that the condition for the existence of a bound state with negative energy $E<0$ differs between these two cases~\cite{1968quantum}. For the spherically symmetric case, decomposing the wave function as $\Psi=\sum_{\ell,m} \f{\psi_\ell(r)}{r}Y_{\ell,m}(\theta,\varphi)$ and focusing on the monopole mode $\ell=0$, one arrives at the form of \eqref{Sch-eq}.  The interior solution is given by a linear combination of sine and cosine functions. Requiring the regularity of $\Psi$ at $r=0$ allows only the odd-parity sine function.  The even-parity solution, which has lower energy in general, is not allowed. As a result, a bound state with negative energy exists if and only if 
$U_0>U_{0,c}$ where the critical value is given by
\be \label{U0csph}
U_{0,{\rm c}}=\f{\pi^2}{4D^2}\,.
\ee
It implies that the square well potential needs to be sufficiently deep and/or wide to satisfy the finite threshold. In contrast, for the planar symmetric case, both of the odd and even functions are allowed, and consequently the bound state with negative energy exists so long as $U_0>0$.  Hence, in the planar symmetric case, the critical value is given by
\be \label{U0cwall}
U_{0,{\rm c}}=0\,.
\ee
Therefore, it is natural to expect that the spontaneous scalarization occurs more easily in the planar symmetric case, compared with the finite threshold of the effective potential required for the spherically symmetric case. We shall show that the analogy indeed applies.

The rest of the paper is organized as follows. In \S \ref{sec:mod} we define the model and notations. Throughout the paper, for simplicity we neglect the mass of the scalar field in vacuum. In \S \ref{sec:sph} we study a spherically symmetric spacetime and clarify that there exists a finite and nonvanishing threshold for the spontaneous scalarization. In \S \ref{sec:wall} we investigate the planar symmetric spacetime and analytically show that the spontaneous scalarization takes place for an arbitrary small absolute value of the tachyonic mass inside a matter source. Then \S \ref{sec:con} is devoted to a conclusion and discussions. In Appendix~\ref{sec:sta} we provide a note on the curvature singularity of the static planar symmetric solution.

\section{The model}
\label{sec:mod}

Let us define the model and summarize the notation that we adopt in this paper. In this section we do not specify the form of the metric nor the matter profile. We work in the Einstein frame and thus consider the Einstein-Hilbert term for the metric $g_{\mu\nu}$ with a canonical kinetic term for a scalar field $\phi$, 
\be \label{action}
S = \int d^4x\sqrt{-g} \mk{\f{R}{16\pi G} - \f{1}{2} \partial_\mu \phi \partial^\mu \phi - V(\phi)} + S_m (\tilde g_{\mu\nu}, \psi_m)\,,
\ee
where the matter action $S_m$ has a nonminimal coupling to the scalar field via the Jordan frame metric $\tilde g_{\mu\nu} \equiv \Omega^2(\phi)g_{\mu\nu}$. 
The Einstein and Klein-Gordon equations are given by
\begin{align} \label{ekgeqs}
G^{\mu\nu} &= 8\pi G (\Omega^2T^{\mu\nu}+T^{(\phi)\mu\nu})\,, \notag\\
\Box \phi &= \f{\pa V_{\rm eff}}{\pa \phi}\,,
\end{align}
where the stress energy tensor for the scalar field and the effective potential are given by 
\begin{align}
T^{(\phi)\mu\nu}&\equiv - \mk{ \f{1}{2} \pa_\lambda\phi\pa^\lambda\phi + V(\phi) } g^{\mu\nu} + \pa^\mu\phi\pa^\nu\phi  ,\notag\\
V_{\rm eff}(\phi)&\equiv V(\phi) - \f{1}{4} \Omega^4(\phi)\tilde T\,,
\end{align}
and the matter stress energy tensors are $T^{\mu\nu} \equiv \f{2}{\sqrt{- g}} \f{\delta S_m}{\delta g_{\mu\nu}}$ and $\tilde T^{\mu\nu} \equiv \f{2}{\sqrt{-\tilde g}} \f{\delta S_m}{\delta \tilde g_{\mu\nu}}$, and the trace is $\tilde T\equiv \tilde g_{\mu\nu}\tilde T^{\mu\nu}$.

We focus on the background solution with $\phi = \phi_0 = $~const., assuming that $\Omega(\phi_0) = 1$, $\Omega_{,\phi}(\phi_0) = 0$ and $V(\phi_0) = V_{,\phi}(\phi_0) = 0$. These conditions are sufficient to guarantee the existence of the GR solution with $\phi=\phi_0$ (see also \cite{Motohashi:2018wdq} for the conditions for more general higher-derivative theory but without nonminimal coupling to matter). Indeed, at $\phi=\phi_0$, the Einstein equation~\eqref{ekgeqs} is simply given by
\be
G^{\mu\nu} = 8\pi G T^{\mu\nu}\,.
\ee
From \eqref{ekgeqs} we see that at
the linear order of the scalar perturbation $\delta\phi = \phi-\phi_0$, the Einstein equation is unchanged from GR. Therefore any GR solution solves it up to the linear order of $\delta \phi$.

On the other hand, the Klein-Gordon equation~\eqref{ekgeqs} for $\phi$ shows that, due  
to the nonminimal coupling, the dynamics of the scalar field is affected by the matter configuration. At the background level, the equation of motion is trivially satisfied by $\phi=\phi_0$.

Let us consider a small perturbation of the scalar field $\delta\phi=\phi-\phi_0$. Since its backreaction to the spacetime geometry and matter is absent at the linear order as mentioned above, we consider the evolution of the scalar perturbation on the fixed GR background metric~\footnote{
This approach is sometimes called the ``decoupling limit'' analysis in the literature. There is no backreaction of the scalar perturbation to the metric and matter at linear order for the canonical model of the spontaneous scalarization~\cite{Harada:1997mr}. For example, in \cite{Cardoso:2013opa}, the spontaneous scalarization of black hole surrounded by matter in the model~\eqref{action} was considered.
To study the stability analytically, the backreaction of the scalar field to spacetime geometry and matter was first neglected, a sufficient condition~\cite{Buell:1995} to develop the instability was derived, and then full numerical analysis was provided. A similar analysis was also employed in \cite{Silva:2017uqg} to analytically study the spontaneous scalarization in the Einstein-scalar-Gauss-Bonnet gravity.}. The perturbed equation of motion is given by 
\be \label{kgpt}
\Box \delta\phi = m^2_{\rm eff} \delta\phi\,,
\ee
with the effective mass 
\begin{align} 
m^2_{\rm eff} &\equiv 
V_{,\phi\phi}(\phi_0) - \Omega_{,\phi\phi}(\phi_0) T\,,
\end{align}
where $T \equiv g_{\mu\nu}T^{\mu\nu}$ and we used $\Omega(\phi_0)=1$ and $\Omega_{,\phi}(\phi_0)=0$.

The instability of $\delta \phi$ amounts to the spontaneous scalarization. For instance, let us suppose a homogeneous and isotropic perfect fluid with $T= -\rho+3P$, and define a bare mass squared $m^2\equiv V_{,\phi\phi}(\phi_0)$, i.e.\ the contribution of the potential to the effective mass squared, and a dimensionless parameter~\footnote{Here we use a notation slightly different from the one used in the literature, i.e.\ $\alpha (\phi) \equiv \f{d \ln \Omega(\phi)}{d\phi}$, $\alpha_0\equiv \alpha(\phi_0)$, and $\beta_0\equiv \alpha_{,\phi}(\phi_0)$, for which with $\Omega(\phi_0)=1$ and $\alpha_0=0$, one obtains $\beta_0=\Omega_{,\phi\phi}(\phi_0)$.} $\hb \equiv 3\Mpl^2 \Omega_{,\phi\phi}(\phi_0)$ measuring the size of the contribution of the conformal factor to the effective mass squared in the unit of the reduced Planck mass $\Mpl\equiv(8\pi G)^{-1/2}$ (up to the overall factor of $3$). We then obtain
\be
m^2_{\rm eff} = m^2 + \f{\hb}{3\Mpl^2} (\rho-3P)\,.
\ee
The case with $m=0$ and $\hb<0$ amounts to the spontaneous scalarization when $\rho-3P>0$. With non-negligible density, $\delta \phi$ exhibits a tachyon instability. It implies that the GR solution is unstable and the system is spontaneously scalarized. While the mass $m$ of the potential is phenomenologically important, it is not crucial to include it to see the shape dependence, which is the main purpose of the present paper. Therefore, for simplicity we assume $m^2\ll |\hb (\rho-3P)/(3\Mpl^2)|$ and focus on the case where the stability is solely determined by the effective mass originated from the matter coupling via the Jordan frame metric. In the following sections we shall therefore set $m=0$.

In \S \ref{sec:sph} and \ref{sec:wall}, we shall consider background GR solutions for two different matter distributions, a spherically symmetric profile and a planar symmetric profile, and highlight a crucial difference between them for the stability of the perturbation $\delta\phi$.

\section{Spherically symmetric spacetime}
\label{sec:sph}

Let us begin with the spherically symmetric spacetime. While we shall focus on the static case, first we write down the metric in a time dependent form
\be \label{bgtsph}
ds^2 = - A(r)dt^2 + \f{F(t)}{B(r)}dr^2 + r^2(d\theta^2+\sin^2\theta d\varphi^2)\,,
\ee
just to be parallel with the structure of \S \ref{sec:wall}. We compute the components of the Einstein tensor and assume a diagonal form of the stress-energy tensor of the matter source. We then find that the Einstein tensor has a nontrivial off-diagonal component 
\be
G^t_r = -\f{\dot F}{rAF} = 0\,.
\ee
Therefore, the only possible solution is $F(t)=1$, and a background solution of the form~\eqref{bgtsph} with nontrivial time dependency does not exist. Therefore, we shall focus on the static case from now on.

\subsection{Background solution}

\begin{figure}[t]
	\centering
	\includegraphics[width=0.45\columnwidth]{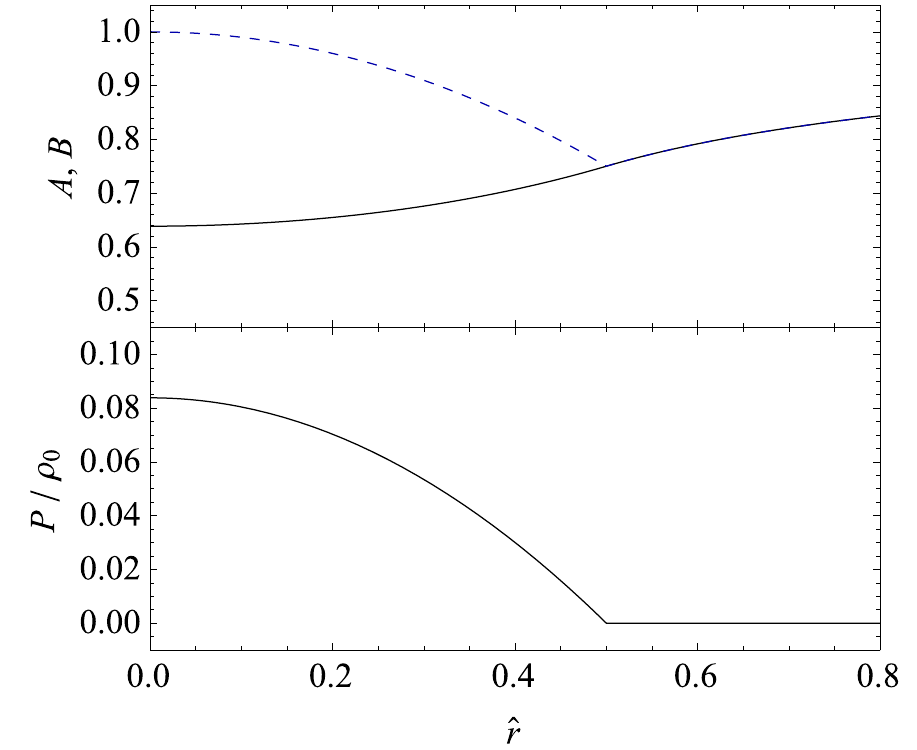}
	\caption{The background solution represented by $A$ (top, black solid line), $B$ (top, blue dashed line), and $P$ (bottom) for the constant density sphere profile with radius $\hat d=0.5$. 
	}
	\label{fig:bg-sph}
\end{figure}

Let us focus on the static, spherically symmetric spacetime given by
\be \label{bg1s}
ds^2 = - A(r)dt^2 + \f{1}{B(r)} dr^2 + r^2(d\theta^2+\sin^2\theta d\varphi^2)\,.
\ee
The Einstein equation is given by
\begin{align} 
G^t_t &= \f{(rB)'}{r^2} -\f{1}{r^2} = -8\pi G\rho\,,\notag\\
G^r_r&= \f{B(rA)'}{r^2A} -\f{1}{r^2} = 8\pi GP\,,
\end{align}
where $\rho(r)$ is the energy density and $P(r)$ is the radial pressure. 
For later convenience, from the Einstein equation one can derive 
\be \label{dABs}
\f{(AB)'}{rA} = \f{2(1-B)}{r^2} - 8\pi G (\rho-P)\,.
\ee

As the simplest example, we consider the density profile with a step function
\be \label{stepspherical}
\rho(r)=
\begin{cases}
\rho_0\,, \quad (0<r<d)\,, \\
0\,, \quad (r>d)\,.
\end{cases}
\ee
It is well known that this system has an exact solution for interior and exterior regions. Let us define $L\equiv (8\pi G\rho_0/3)^{-1/2}$ corresponding to the Jeans length and normalize the variables as $\hat t\equiv t/L$ and $\hat r\equiv r/L$. With the normalized variables, the interior and exterior Schwarzschild solutions are written as
\begin{align} 
 A &= \f{1}{4} \mk{ 3\sqrt{1-\hat d^2} - \sqrt{1-\hat r^2} }^2\,,
\quad
B = 1 - \hat r^2\,, \quad (0<\hat r< \hat d)\,, \notag\\
A&=B = 1- \f{\hat d^3}{\hat r}\,,
 \quad (\hat r> \hat d)\,,
\end{align}
where $\hat d = d/L$. The pressure is given by analytically integrating the Tolman-Oppenheimer-Volkoff (TOV) equation as
\be
P = 
\begin{cases} \displaystyle
\rho_0\f{ \sqrt{1-\hat r^2} - \sqrt{1-\hat d^2} }{ 3\sqrt{1-\hat d^2} - \sqrt{1-\hat r^2} }\,,  &(0<\hat r<\hat d)\,,\\
0\,, &(\hat r>\hat d)\,,
\end{cases}
\ee
which has the maximum value at $r=0$, monotonically decreases as $r$ increases, and reaches zero at $r=d$. This background solution is shown in Fig.~\ref{fig:bg-sph} for the case $\hat d=0.5$. Note that the normalized Schwarzschild radius is not $\hat d$ but $\hat d^3$. For $0<\hat r<\hat d$, $A(\hat r)=0$ corresponds to
\be
\hat r^2=9\hat d^2 -8\,.
\ee
Therefore, so long as we consider 
\be
\hat d<\f{2\sqrt{2}}{3}\approx 0.943\,,
\ee
$A(\hat r)$ is positive and does not cross zero for $\hat r>0$.

\subsection{Perturbation}

Next, we study perturbation about the background analytic solution. As mentioned above, the metric and matter remain as the GR solution at linear order of the scalar perturbation $\delta\phi = \phi-\phi_0$. Hence we focus on the perturbed Klein-Gordon equation on the fixed GR metric solution. By decomposing the perturbation into spherical harmonics as
\be \delta\phi= \sum_{\ell, m}\f{\sigma_{\ell }(t,r)}{r}Y_{\ell m}(\theta,\varphi) , \ee
from \eqref{kgpt} we can write down the evolution equation as
\be
-\f{\pa^2\sigma_\ell}{\pa\hat t^2} 
+ \f{\pa^2\sigma_\ell}{\pa \hat r_*^2} 
= A \mk{ \f{1-B}{\hat r^2} 
- \f{3}{2} \f{\rho-P}{\rho_0}   
+ \f{\ell(\ell+1)}{\hat r^2} 
+ L^2m_{\rm eff}^2
} \sigma_\ell\,,
\ee
where 
$\hat r_*$ is the normalized tortoise coordinate defined via $d\hat r/d\hat r_*=\sqrt{AB}$. We also used \eqref{dABs} to simplify the right-hand side.

The relation between $\hat r$ and $\hat r_*$ can be analytically obtained as
\be
\hat r_* (\hat r) = 
\begin{cases}
\displaystyle \f{2}{\sqrt{8 - 9 \hat d^2} } \arctan \mk{\f{ \hat r\sqrt{8 - 9 \hat d^2} }{3 \sqrt{(1 - \hat d^2)(1 - \hat r^2)} - 1} }\,, & (0<\hat r<\hat d)\,, \\
\displaystyle \hat r - \hat d + \hat d^3 \log\mk{ \f{\hat r-\hat d^3}{\hat d-\hat d^3} } 
+ \f{2}{\sqrt{8 - 9 \hat d^2} } \arctan\mk{ \f{\hat d \sqrt{8 - 9 \hat d^2}}{2 - 3 \hat d^2} }\,, & (\hat r>\hat d)\,.
\end{cases}
\ee
We can also obtain the inverse function as
\be \label{rrstar}
\hat r (\hat r_*)=
\begin{cases}
\displaystyle \f{\sqrt{8 - 9 \hat d^2} \sin(\hat r_* \sqrt{8 - 9 \hat d^2}/2)}{\cos (\hat r_* \sqrt{8 - 9 \hat d^2}/2) + 3 \sqrt{1 - \hat d^2}}\,, & (0<\hat r_*<\hat d_*)\,, \\
\displaystyle \hat d^3 \kk{ 1 + W \mk{ (\hat d^{-2}-1) \exp \kk{\hat d^{-2}-1+\f{\hat r_*}{\hat d^{3} } - \f{2}{\hat d^{3}\sqrt{8 - 9 \hat d^2} } \arctan \mk{ \f{\hat d \sqrt{8 - 9 \hat d^2}}{2 - 3 \hat d^2} }  }  }}\,, & (\hat r_*>\hat d_*)\,,
\end{cases}
\ee
where $\hat d_*\equiv \hat r_*(\hat d)$ and $W(z)$ is the product logarithm or the Lambert function satisfying $z=We^{W}$. Note that $\lim_{\hat r\to 0} \hat r_* = 0$ and $\lim_{\hat r\to \infty} \hat r_* = \infty$, and $\hat r_* \approx \hat r$ at the leading order of the approximation $\hat d\ll 1$.

Below we set $m^2\equiv V_{,\phi\phi}(\phi_0)=0$ and focus on the $\ell=0$ monopole mode as it is the most unstable mode. Using the definition $\Omega_{,\phi\phi}(\phi_0)\equiv \hb/(3\Mpl^2)$, we obtain
\be \label{sigmaeq-sph}
-\f{\pa^2\sigma_0}{\pa \hat t^2} 
+ \f{\pa^2\sigma_0}{\pa \hat r_*^2} 
= A \kk{ \f{1-B}{\hat r^2} 
+ \mk{ \hb - \f{3}{2} } \f{\rho-P}{\rho_0}
} \sigma_0\,.
\ee 
By using the Fourier decomposition $\sigma_0 = \int \f{d\omega}{2\pi} e^{-i\omega \hat t} \psi(\hat r_*)$, we obtain the Scr\"odinger-type equation
\be
\mk{- \f{d^2}{d\hat r_*^2} + U } \psi
= E \psi\,,
\ee
with the effective potential and energy as
\be
U(\hat r_*) \equiv A \kk{ \f{1-B}{\hat r^2} 
+ \mk{ \hb - \f{3}{2} } \f{\rho-P}{\rho_0}
}\,, \quad
E \equiv \omega^2\,,
\ee
where the right-hand side of the effective potential $U(\hat r_*)$ is understood as a function of $\hat r_*$ after substituting $\hat r = \hat r (\hat r_*)$ given in \eqref{rrstar}. Since $\delta\phi\sim \psi(\hat r_*)/\hat r$ and $\lim_{\hat r\to 0} \hat r_* = 0$, the regularity of $\delta\phi$ at the origin $\hat r=0$ requires the boundary condition $\psi(0)=0$. We note that the existence of a bound state satisfying $|\psi(\pm\infty)|<\infty$ with negative energy $\omega^2<0$ amounts to the instability as the perturbation $\delta\phi$ grows up exponentially.

\begin{figure}[t]
	\centering
	\includegraphics[width=0.45\columnwidth]{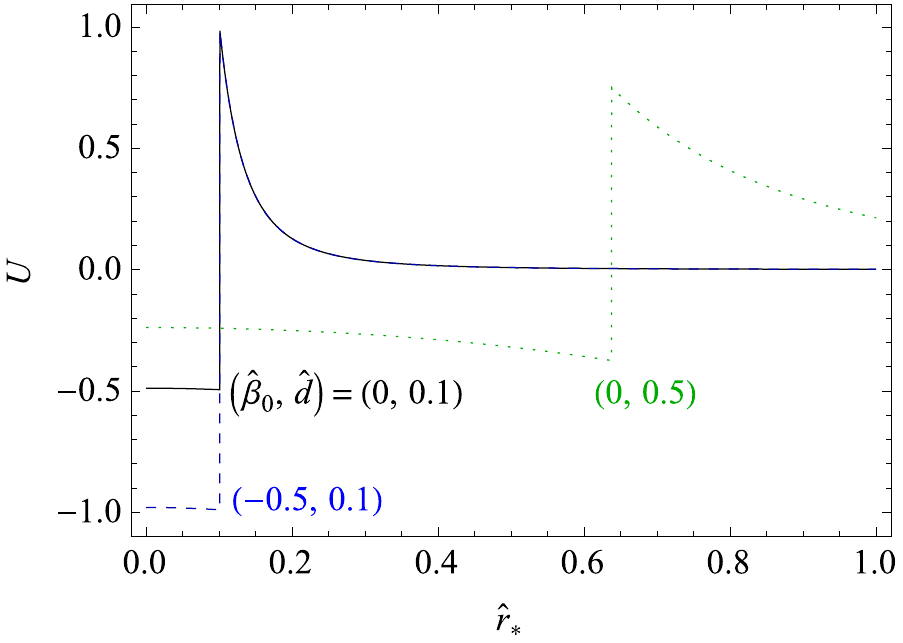}
	\includegraphics[width=0.45\columnwidth]{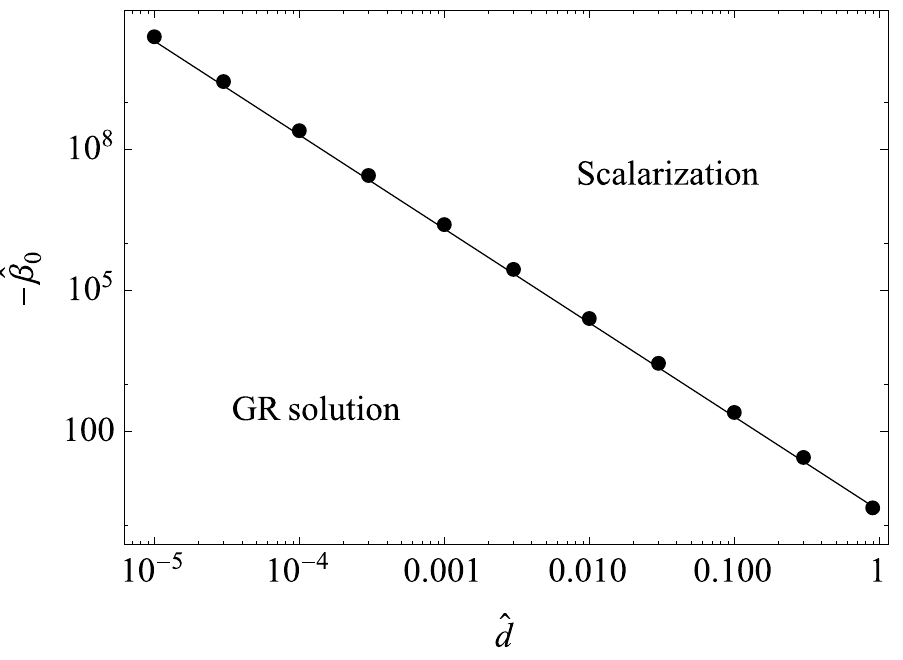}
	\caption{The potential $U(\hat r_*)$ (left) 
	and the critical value $\hbc$ (right) for the spherically symmetric step function density profile \eqref{stepspherical}. 
	}
	\label{fig:Uath-sph}
\end{figure}

The potential $U(\hat r_*)$ is shown in Fig.~\ref{fig:Uath-sph}, which is determined by two dimensionless parameters, $\hat d$ and $\hb$. Since we are interested in the spontaneous scalarization, we focus on the case $\hb<0$. The larger $|\hb|$ is, the deeper the depth of the well is, implying that it is easier to have a negative energy bound state. On the other hand, as $\hat d$ increases, the width of the negative region increases.

To check the existence of the bound state with negative energy, we can exploit a theorem of the Strum-Liouville theory on the number of nodes of eigenfunctions~\cite{Courant1} (for a related method, see \cite{Kimura:2017uor,Kimura:2018eiv}): if eigenfunctions $\psi_0,\psi_1,\cdots$ are ordered according to increasing eigenvalues with $E_0<E_1<\cdots$, the $n$-th eigenfunction has $n$ nodes. Hence, solving the Schr\"odinger equation \eqref{Sch-w} with $E=0$, we can check whether it is the ground state, i.e.\ the bound state with the lowest $E$, by looking at whether it does not or does have nodes for $\hat r_*>0$. Note that from the boundary condition, the solution always has a node at $\hat r_*=0$. However, it corresponds to the existence of the even-parity solution with negative energy, which, however, is prohibited by the requirement of the regularity of $\delta\phi$ at the origin.  Therefore, we focus on nodes for $\hat r_*>0$.  In particular, we are interested in clarifying the critical value $\hbc$ such that the zero energy Schr\"odinger equation starts to have a node for $\hb<\hbc$. It means that for $\hb \geq \hbc$ the zero energy bound state is the ground state without any nodes for $\hat r_*>0$. Therefore, in practice, for a given $\hat d$ we solve the zero energy Sch\"odinger equation with the odd-parity boundary condition $\psi(0)=0$ and $\psi'(0)=1$ with a test value of negative $\hb$ sufficiently close to $0$ so that the solution does not have a node for $\hat r_*>0$, and we iterate the calculation with a smaller $\hb$ until we identify the critical value $\hbc$ below which the solution starts to have a node for $\hat r_*>0$. Then, from the nodal theorem, a node of the zero energy bound state guarantees the existence of a negative energy bound state, which leads to the instability of $\delta \phi$, i.e.\ the spontaneous scalarization.

The numerically obtained $\hbc$ for various $\hat d$ is depicted in Fig.~\ref{fig:Uath-sph}. The dots from 
$(\hat d,-\hat\beta_0) = (10^{-5},2.47 \times 10^{10} )$ to $(0.9,2.29)$
denote numerically calculated values of the critical value $\hbc$, 
which can be well approximated by
\be \label{b0csph} -\hbc \approx \f{2}{\hat d^2}, \ee 
shown as the solid line in Fig.~\ref{fig:Uath-sph}. 
The scaling of \eqref{b0csph} precisely matches the physical intuition from \eqref{U0csph}.
It can also be understood from an order estimation following from the dispersion relation $\omega^2\sim d^{-2} - |m^2_{\rm eff}| \sim \rho_0(\hat d^{-2} - |\hat\beta_0|)$, which needs to be negative for instability. The result also matches the expectation from the shape of the potential. For a fixed $\hat d$, as $\hb$ decreases and subceeds the critical value $\hbc$, the spontaneous scalarization occurs. This is because a deeper potential allows the existence of a bound state with negative energy, leading to the instability that indicates the onset of scalarization. On the other hand, for a fixed $\hb$, as $\hat d$ decreases, the spontaneous scalarization requires larger $|\hat\beta_0|$. Physically, the spontaneous scalarization requires a sufficiently large radius, dense profile, or large tachyon mass due to the conformal factor.

\section{Planar symmetric spacetime}
\label{sec:wall}

Let us see how the results in \S\ref{sec:sph} are changed if we consider a matter profile with a different morphology. As an extreme example, we consider a planar symmetric matter configuration, 
and correspondingly a planar symmetric spacetime whose metric is given by 
\be \label{bgt}
ds^2 = - a^2(z)dt^2 + f^2(t)b^2(z)(dx^2+dy^2) + dz^2\,.
\ee
Writing down all the components of the Einstein tensor and supposing a diagonal form of the stress-energy tensor of the matter component, we obtain a condition from $(t,z)$ component
\be
G^t_z = -\f{2\dot f}{a^2f} \mk{\ln \f{a}{b} } ' = 0\,,
\ee
where a prime denotes a derivative with respect to $z$. It suggests the two possible branches, $f(t)=1$ or $b(z)=a(z)$, where we absorbed the proportionality constant in a redefinition of $b(z)$ or $x,y$, respectively. However, as shown in Appendix~\ref{sec:sta}, the static branch $f(t)=1$ with the step function density profile ends up with a curvature singularity. While it may be possible to consider other density profiles to derive a solution without the curvature singularity, the fact that the simplest profile is plagued with the curvature singularity is not attractive. Hence we focus on the second branch with $b(z)=a(z)$: 
\be \label{bgt2}
ds^2 = - a^2(z)dt^2 + f^2(t)a^2(z)(dx^2+dy^2) + dz^2\,.
\ee

Outside the matter source, using the vacuum Einstein equation, one can confirm that the contraction of the Weyl tensor identically vanishes on vacuum, implying that there is no curvature singularity for the metric~\eqref{bgt2}. Furthermore, one can also check that all the components of the Riemann tensor are identically vanishing, which implies that the vacuum solution of the form \eqref{bgt} is locally equivalent to the Minkowski metric through a coordinate transformation.
The $(t,t)$ component of the vacuum Einstein equation is given by
\be
G^t_t = -\f{\dot f^2}{a^2f^2} + \f{a'^2}{a^2} + \f{2a''}{a} = 0\,.
\ee 
In what follows, we consider the case where the separation of variables applies, i.e.\ the case with $f(t) = e^{Ht}$, where $H$ is a positive constant, not only outside but also inside the matter source.

\subsection{Background}

Let us consider the metric 
\be
ds^2 = - a^2(z)dt^2 + e^{2Ht}a^2(z)(dx^2+dy^2) + dz^2\,,
\ee
with the planar and $Z_2$ symmetric matter configuration. Here, by the $Z_2$ symmetry, we mean that the configuration is invariant under the $z$-parity, i.e.\ $z\to -z$. A solution for a similar setup of the thick domain wall was derived in \cite{Bonjour:1999kz}, where the domain wall is formed by a scalar field. 
See also \cite{Ipser:1983db} for the most general reflection-symmetric solution to Einstein equations for a planar domain wall.
In the present case we simply assume a matter profile which respects the planar and $Z_2$ symmetries. Considering an anisotropic perfect fluid, the stress-energy tensor is given by
\be
T^{\mu\nu} = (\rho+P_x)u^\mu u^\mu + P_x g^{\mu\nu} + (P_z-P_x)z^\mu z^\nu\,,
\ee
where $u^{\mu} = (1/a)(\partial / \partial t)^{\mu}$, $z^{\mu} = (\partial / \partial z)^{\mu}$, and we have used $P_y=P_x$. The Einstein equation is then given by
\begin{align} \label{Eineq}
G^t_t&= \f{a'^2-H^2}{a^2} + \f{2a''}{a} = -8\pi G \rho\,,\notag\\
G^x_x=G^y_y&= \f{a'^2-H^2}{a^2} + \f{2a''}{a} = 8\pi G P_x\,,\notag\\
G^z_z&= \f{3(a'^2-H^2)}{a^2} = 8\pi G P_z\,.
\end{align}
From \eqref{Eineq} we immediately see that $P_x=-\rho$. Therefore the matter stress-energy tensor is specified by $\rho$ and $P_z$.

Given $\rho(z)$, one can solve \eqref{Eineq} for $a$ and $P_z$. For the following we assume that $a>0$ and $3\rho+P_z \geq 0$, the latter of which is weaker than the null energy condition $\rho+P_z \geq 0$ so long as $\rho\geq 0$. We further assume that $(3\rho+P_z )|_{z=0} > 0$. For general $\rho$ and $P_z$, we can derive
\be
\f{a''}{a} = -4\pi G (3\rho+P_z)\,.
\ee
Hence, under the condition $3\rho+P_z \geq 0$ it holds that $a''(z)\leq 0$. It also follows that $a''(0) < 0$. Since the matter configuration respects the planar and $Z_2$ symmetries, $a(z)$ is also an even function. Thus we impose the condition $a(0)=1$ and $a'(0)=0$ at $z=0$, and focus on the positive region $z>0$ as the negative region $z<0$ is given by a reflection of the positive region. Combining $a''(z) \leq 0$ for $z>0$, $a''(0) < 0$ and $a'(0)=0$, we obtain $a'(z)<0$; i.e.\ $a(z)$ is monotonically decreasing, for $z>0$. In particular, if there exists a boundary $z=\pm d$ of matter such that $\rho=P_z=0$ for $|z| \geq d$, it holds that $a'(d)<0$, which can be used as a boundary condition to connect to the vacuum solution.

Using the boundary condition of $a'(d)<0$, we can solve the $(z,z)$ component of the vacuum Einstein equation to obtain the exterior solution
\be \label{vaca} a(z)= 
\begin{cases}
H (z + z_h)\,, & (z < - d)\,, \\
- H (z - z_h)\,, & (z > d)\,,
\end{cases}
\ee
where $z_h$ is an integration constant which can be fixed by matching the values of $a(d)$ between the exterior solution and the interior solution. We note that $a(z)=0$ at $z=\pm z_h$, which defines the cosmological horizon~\cite{Bonjour:1999kz}.

The interior solution for $a(z)$ and the pressure $P_z(z)$ can be obtained for a given matter density distribution $\rho$ by solving the Einstein equation.
To compare with the case of a sphere profile with constant density in \S\ref{sec:sph}, 
we consider a wall with constant density extended for $-d<z<d$, 
\be \label{rho-wall}
\rho(z) = \begin{cases}
\rho_0\,, & (|z|<d)\,, \\
0\,, & (|z|>d)\,.
	  \end{cases}
\ee
By introducing the Jeans scale $L\equiv (8\pi G \rho_0/3)^{-1/2}$, in the following we work with the normalized variables $\hat t \equiv t/L$, $\hat z \equiv z/L$, and $\gamma\equiv HL$, which is the ratio between the Jeans scale and the cosmological horizon scale. Note also that $\gamma^2=3H^2/(8\pi G\rho_0)=1/\Omega_{\rm w}$ where $\Omega_{\rm w}$ is the analogy of the density parameter of the wall. In the following we denote $a'=da/d\hat z$.

The interior solution for the matter density profile~\eqref{rho-wall} can be obtained by solving the Einstein equation~\eqref{Eineq}.
With the normalized variables, it can be rewritten as 
\begin{align}
\label{aeq}  &2a\f{d^2a}{d\hat z^2} + \mk{\f{da}{d\hat z}}^2 + 3 a^2 - \gamma^2 = 0\,, \\
\label{Pzeq} &\f{P_z}{\rho_0} 
= \f{1}{a^2} \kk{ \mk{\f{da}{d\hat z}}^2-\gamma^2 }\,.
\end{align}

\begin{figure}[t]
	\centering
	\includegraphics[width=0.45\columnwidth]{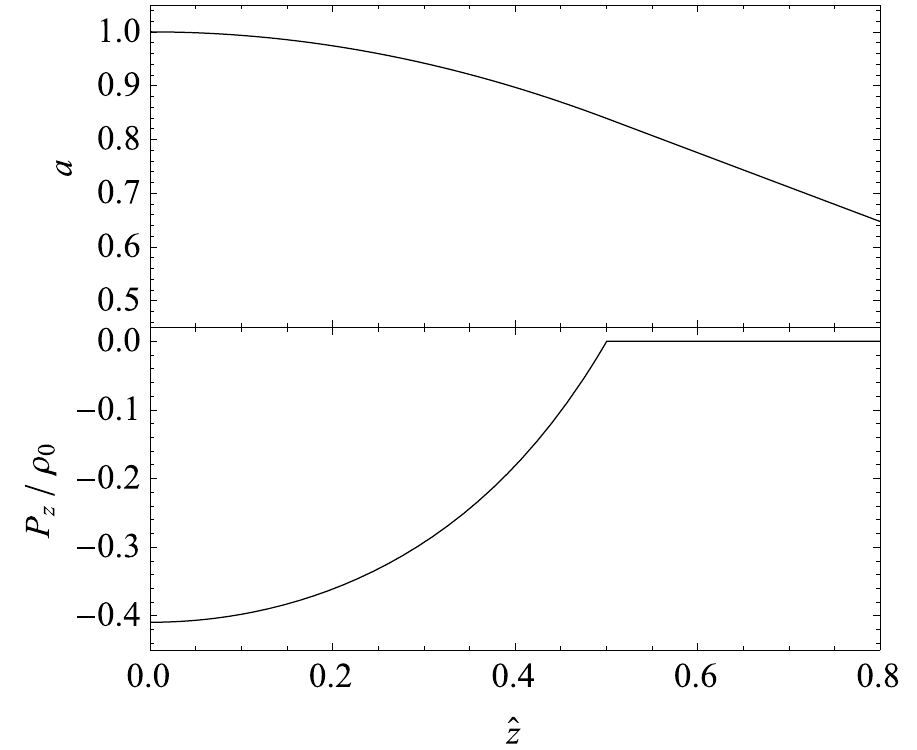}
	\caption{The background solutions $a$ (top) and $P_z$ (bottom) for the constant density wall profile \eqref{rho-wall} with $\hat d=0.5$. 
	}
	\label{fig:bg-wall}
\end{figure}

Let us impose the condition $\rho_0+P_z(0)> 0$, which is necessary for the null energy condition and sufficient for the condition $3\rho+P_z>0$ considered above.
From $P_z(0)/\rho_0=-\gamma^2$ we obtain the condition $\gamma< 1$.
For the parameter region $0<\gamma< 1$, we numerically solved \eqref{aeq} for the interior solution of $a(z)$ with the boundary condition $a(0)=1$ and $a'(0)=0$, and we obtain $P_z(z)$ from \eqref{Pzeq}.
We then confirmed that there exists the boundary of the wall $\hat z=\hat d$ where $P_z(\hat d)=0$.
The parameter region $0<\gamma< 1$ corresponds to $0<\hat d<\hat d|_{\gamma=1}\approx 1.57$.
The example of the case $\hat d=0.5$ is depicted in Fig.~\ref{fig:bg-wall}.
We also confirmed that $\rho_0+P_z(\hat z)> 0$ is satisfied for $0<\hat z< \hat d$.
Since the condition $3\rho+P_z>0$ is satisfied, $a(\hat z)$ is monotonically decreasing for $0<\hat z<\hat d$, and connects to the exterior vacuum solution~\eqref{vaca}.

\subsection{Perturbation}

\begin{figure}[t]
	\centering
	\includegraphics[width=0.45\columnwidth]{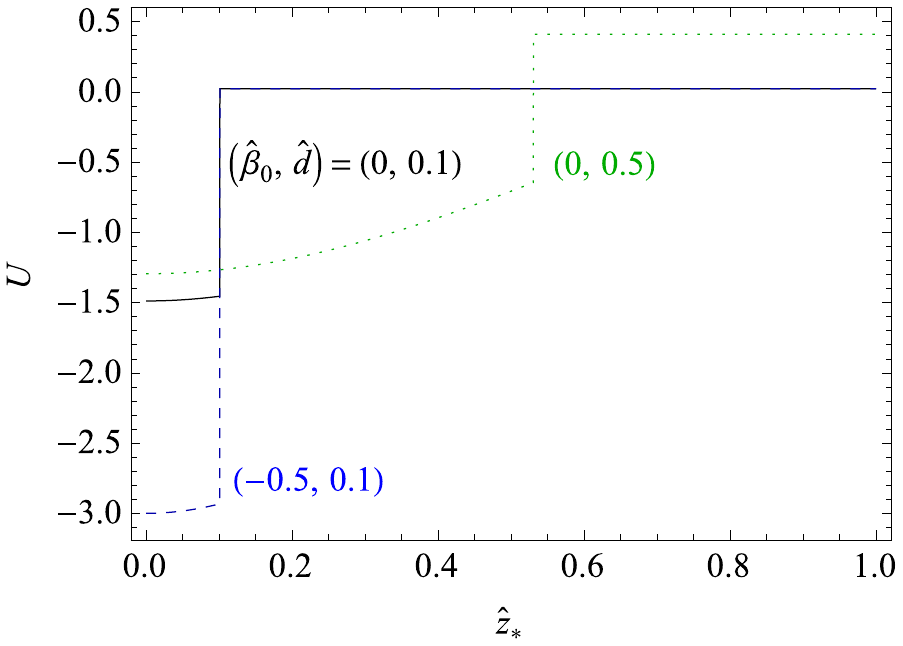}
	\includegraphics[width=0.45\columnwidth]{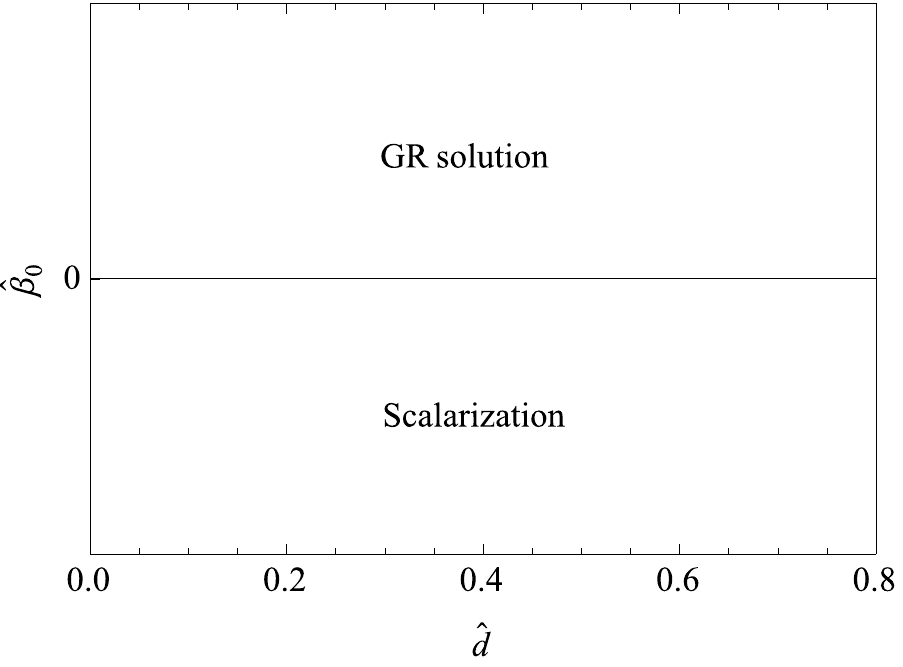}
	\caption{The potential $U(\hat z_*)$ (left) for the constant density wall profile \eqref{rho-wall} and the critical value $\hbc$ (right) for the planar symmetric step function density profile. 
	}
	\label{fig:Uath-wall}
\end{figure}

Next let us consider the scalar perturbation $\delta\phi=\phi-\phi_0$ on the above background GR solution, since the metric and matter remain unchanged at the linear order of $\delta\phi$.
Unlike the spherically symmetric case in \S\ref{sec:sph}, it is possible to show that $\hbc=0$ for a general profile 
\be
\label{rho-general} 
\rho(z) = \begin{cases}
\rho_{\rm int} (z)\,, & (|z|<d)\,, \\
0\,, & (|z|>d\,),
\end{cases}
\ee
respecting the $Z_2$ and planar symmetries with the normalization $\int_{-d}^d dz \rho_{\rm int}(z) = 2\rho_0 d$. The density profile can be originated from an external source, a scalar field, or curvature invariants. Below we consider the general profile~\eqref{rho-general}, and use the constant density profile~\eqref{rho-wall} only for presentation of a concrete effective potential in Fig.~\ref{fig:Uath-wall}.

From \eqref{kgpt} we can derive the equation of motion for perturbation of the scalar field.
By decomposing the perturbation into Fourier mode as 
\be
\delta\phi = \int \f{d^2k}{(2\pi)^2}\f{\sigma_{\bf k}(\hat t,\hat z)}{e^{\gamma \hat t}a} e^{i(k_xx+k_yy)}\,,
\ee
with ${\bf k}=(k_x,k_y)$, the evolution equation takes the form of
\be
- \f{\pa^2\sigma_{\bf k}}{\pa\hat t^2} + \f{\pa^2\sigma_{\bf k}}{\pa \hat z_*^2} - a^2 \kk{ L^2 m_{\rm eff}^2 + \f{L^2 k^2}{e^{2\gamma\hat t}a^2} + \f{3}{2} \f{1}{\rho_0}\mk{\f{P_z}{3}-\rho}  } \sigma_{\bf k} = 0\,,
\ee
where $k^2\equiv k_x^2+k_y^2$ and $\hat z_*$ is the normalized tortoise coordinate defined via $d\hat z/d\hat z_*=a(\hat z)$.

In parallel to the spherically symmetric case in \S\ref{sec:sph}, we focus on the $k_x=k_y=0$ mode as it is the most unstable mode, set $m^2=0$, and use the notation $\hb\equiv 3\Mpl^2\Omega_{,\phi\phi}(\phi_0)$. We also note that the trace of the stress-energy tensor is given by $T=-\rho+2P_x+P_z=-3\rho+P_z$ in the present case. By using the Fourier decomposition $\sigma_{\bf 0}=\int \f{d\omega}{2\pi} e^{-i\omega \hat t} \psi(\hat z_*)$, we obtain
\be
\kk{- \f{d^2}{d \hat z_*^2} + 3\mk{ -\hb + \f{1}{2} } \f{a^2}{\rho_0}\mk{\f{P_z}{3}-\rho} }\psi = \omega^2 \psi\,.
\ee 
Here, unlike the spherically symmetric case in \S\ref{sec:sph}, we need to take into account the fact that $\delta \phi \sim e^{(-i\omega -\gamma )\hat t} \psi(\hat z_*)/a(\hat z_*) = e^{(|\omega| -\gamma )\hat t} \psi(\hat z_*)/a(\hat z_*)$ where we consider pure imaginary frequency $\omega = i|\omega|$. This means that the instability or exponential growth of $\delta\phi$ corresponds to $|\omega|>\gamma$. Therefore, it is more appropriate to work in the notation 
\be \label{Sch-w}
\mk{- \f{d^2}{d \hat z_*^2} + U }\psi = E \psi\,
\ee
with the potential and energy
\be 
U(\hat z_*) \equiv 3\mk{ - \hb + \f{1}{2} } \f{a^2}{\rho_0}\mk{\f{P_z}{3}-\rho} + \gamma^2 , \qquad
E\equiv - \omega^2 + \gamma^2\,.
\ee
The right-hand side of the potential $U$ is understood as a function of $\hat z_*$ after substituting the numerical solution $\hat z=\hat z(\hat z_*)$. The exponential growth of $\delta\phi$ amounts to the bound state with negative energy $E<0$.

As a demonstration, the potential $U(\hat z_*)$ for the constant density wall profile \eqref{rho-wall} is depicted in Fig.~\ref{fig:Uath-wall}. While in practice we set the model parameter $\gamma$ first and determine $\hat d$ for each $\gamma$, to highlight the interplay to the results obtained in \S\ref{sec:sph}, we showed the value of $\hat d$. As expected, the depth of the potential is deeper for a larger value of $|\hb|$.

Let us go back to the argument with the general density profile~\eqref{rho-general}.
As mentioned above, the instability or the spontaneous scalarization amounts to the existence of a bound state with negative energy $E<0$. From the theorem of the Strum-Liouville theory, at the critical value $\hb=\hbc$ for the instability, the Schr\"odinger equation~\eqref{Sch-w} with $E=0$ allows a bound state without nodes. Once one considers $\hb<\hbc$, the bound state for $E=0$ starts to have a node, implying the existence of a bound state for $E<0$. Thus, let us focus on the bound state for $E=0$ without nodes.

First, let us focus on the interior solution obeying the Schr\"odinger equation~\eqref{Sch-w} with $E=0$. Since the effective potential has the $Z_2$ symmetry, without loss of generality we can assume that $\psi(\hat z_*)$ is either even or odd. We emphasize here that unlike the spherically symmetric case in \S\ref{sec:sph}, both odd- and an even-parity solutions are allowed for the planar symmetric case.  For the odd-parity case, the solution always has at least one node at $\hat z_*=0$. Therefore, the bound state without a node respects the even-parity, satisfying the boundary condition $\psi(0)=1$ and $\psi'(0)=0$.

Next, let us focus on a possibility of nodes for $\hat z_*>\hat d_*$ for such a even-parity solution. Since $U=\gamma^2=$~const.\ for $\hat z_*>\hat d_*$, the exterior solution is simply given by 
\be
\psi=c_+e^{\gamma \hat z_*} + c_-e^{- \gamma \hat z_*}\,,
\ee
where $c_\pm$ are integration constants. Without loss of generality, we assume that $c_- \ne 0$ since the solution with $c_- = 0$ is either trivial (if $c_+ = 0$) or unbounded (if $c_+\ne 0$). This solution has either one or zero node for $\hat z_* > \hat d_*$, depending on the sign of $c_+/c_-$: the solution with $c_+/c_- > 0$ does not have nodes for $\hat z_* > \hat d_*$. However, since $|\psi|\to \infty$ for $\hat z_*\to\infty$, it is not a bound state. The solution with $c_+=0$ does not have nodes for $\hat z_*>\hat d_*$ and is a bound state as $\psi$ is decaying as $\hat z_*\to\infty$.  The solution with $c_+/c_-<0$ has one node for $\hat z_*>\hat d_*$, and $|\psi|\to \infty$ for $\hat z_*\to\infty$. Therefore, the case $c_+ = 0$ is realized for the critical value $\hb=\hbc$. For a given function $\psi$, we can check if $c_+ = 0$ or not by using the identity 
\be \label{wallcrit}
\f{d\psi}{d\hat z_*}(\hat d_*) + \gamma \psi (\hat d_*) = 2\gamma c_+ e^{\gamma \hat d_*}\,.
\ee
For the critical value $\hb=\hbc$, the left-hand side should vanish.

In summary, the critical value $\hb=\hbc$ is defined by the condition that the Schr\"odinger equation~\eqref{Sch-w} with $E=0$ allows an even-parity bound state without a node. The even parity is guaranteed by the boundary condition $\psi(0)=1$, $\psi'(0)=0$, and the absence of nodes for $\hat z_*>\hat d_*$ is guaranteed by the condition $d\psi/d\hat z_*(\hat d_*) + \gamma \psi (\hat d_*)=0$.

Interestingly enough, under the assumption $a>0$ and $3\rho+P_z>0$ for $|\hat z_*|<\hat d_*$, for any matter profile without any approximation, we can show
\be \label{b0cwall} \hbc =0, \ee 
for the planar symmetric case, in contrast to \eqref{b0csph} for the spherically symmetric case.  It matches the physical intuition from \eqref{U0cwall}.  The proof is as follows.
First, using \eqref{Eineq}, we can show that $\psi=a$ is a solution of the Schr\"odinger equation~\eqref{Sch-w} with $E=0$ and $\hb = 0$. This solution also satisfies the boundary condition $\psi(0)=1$ and $\psi'(0)=0$. The assumption $a>0$ guarantees the absence of nodes. Furthermore, under the assumption $a>0$, $3\rho+P_z \geq 0$ and $(3\rho+P_z)|_{z=0} > 0$, the exterior solution is given by \eqref{vaca}, and hence $da/d\hat z(\hat d)=-\gamma$. Since $\psi=a$, this equation translates to $d\psi/d\hat z_*(\hat d_*) = - \gamma \psi(\hat d_*)$, which is nothing but the desired condition. This proves that $\hbc = 0$ for a wide class of planar symmetric configurations, as shown in Fig.~\ref{fig:Uath-wall} for a particular case.

As a complementary check, we performed numerical calculations and confirmed that $\hbc = 0$ within the numerical error of order $10^{-16}$. In numerical calculations, we solve \eqref{Sch-w} in terms of $\hat z$ rather than $\hat z_*$ to avoid the numerical error caused by the interpolating function $\hat z=\hat z(\hat z_*)$. For errors associated with the numerical integration of two differential equations \eqref{aeq} and \eqref{Sch-w} for the background and the perturbation respectively, we require that the relative errors remain less than $10^{-16}$. Consequently we find out that down to the order of the numerical error, the system exhibits the scalarization for $\hb<-10^{-16}$, whereas the GR solution is stable for $\hb>10^{-16}$. These are consistent with the analytically obtained value of $\hbc=0$.

This result highlights the shape dependence of the spontaneous scalarization explicitly.
For the spherically symmetric profile shown in Fig.~\ref{fig:Uath-sph}, $\hbc\gtrsim O(1)$ for $0<\hat d<0.8$, meaning that the spontaneous scalarization requires a sufficiently large radius, dense profile, or large tachyon mass due to the conformal factor. In contrast, for the planar symmetric profile (such as the one shown in Fig.~\ref{fig:Uath-wall}), the scalarization occurs for $\hb<0$, whereas the GR solution is stable for $\hb>0$. So long as the conformal factor $\Omega$ contributes negatively to the effective mass squared of the scalar field, regardless of the amplitude of the tachyon mass and the density profile $\rho$, the scalarization occurs. This shape dependence matches the physical intuition acquired from the analogous Schr\"odinger problem for the square well potential in the spherically symmetric profile and the planar symmetric profile. As mentioned in \S\ref{sec:intro}, for the former case a bound state with negative energy exists if the depth and width of the well exceed a critical value, whereas in the latter case the critical value is $0$.

\section{Conclusion and discussions}
\label{sec:con}

We have investigated the canonical scalar-tensor theory with coupling to the matter sector via a conformally transformed metric and highlighted how essential the morphology of the matter profile is for the realization of the spontaneous scalarization. The model allows for a GR solution with a constant scalar field profile, which remains unchanged at the linear order of the perturbation of the scalar field. We have studied the stability of the linear perturbation of the scalar field, where the instability amounts to the spontaneous scalarization. Since the evolution equation of the perturbation takes the form of the Schr\"odinger equation with an effective potential depending on the mass due to the conformal factor and the stress energy tensor of matter, the instability/stability of scalar perturbation is translated into the existence/absence of the bound state with negative energy. Using the theorem of the Strum-Liouville theory, the boundary between the existence and the absence of a bound state with negative energy is obtained by simply demanding that there is a bound state of zero energy without nodes, corresponding to the ground state. In this way we derived the critical value $\hbc$ of the normalized mass squared due to the conformal factor, below which the zero energy bound state has a node, namely, the spontaneous scalarization occurs.

We focused on the spherically and planar symmetric profiles of matter configuration, and clarified that the critical value $\hbc$ has sensitive dependency on the morphology of the matter profile. For the spherically symmetric step function profile of the matter density, the background GR solution is given by the interior and exterior Schwarzschild solutions. We numerically solved the perturbation equation and obtained the critical value $\hbc$ shown in Fig.~\ref{fig:Uath-sph} as a function of the radius $\hat d$ of the sphere normalized by the Jeans length. 
The fitting function of the critical values is given by $-\hbc\approx 2\hat d^{-2}$ in \eqref{b0csph}, which precisely matches the physical intuition from \eqref{U0csph}, and can also be understood from the simple estimation of the dispersion relation.
This implies that the spontaneous scalarization requires a sufficiently large radius, dense profile, or large tachyon mass due to the conformal factor. On the other hand, for the planar symmetric profile, for a general density profile that satisfies $a > 0$, $3\rho+P_z \geq 0$ and $(3\rho+P_z)|_{z=0} > 0$, we analytically proved that $\hbc=0$ as in \eqref{b0cwall}, which matches the physical intuition from \eqref{U0cwall}. Therefore, the spontaneous scalarization occurs so long as $\hb<0$ however small the tachyonic mass from the conformal factor is. Our proof applies to general density profile~\eqref{rho-general} respecting the $Z_2$ and planar symmetries, which can be originated from an external source, a scalar field, or curvature invariants such as the Gauss-Bonnet or the Chern-Simons terms. These results explicitly clarify the shape dependence of the spontaneous scalarization. As already mentioned in \S\ref{sec:intro}, our result matches the physical intuition acquired from the analogous simple Schr\"odinger problem with the square well potential in the spherically symmetric and planar symmetric setups.

The models of the spontaneous scalarization have been extensively explored for spherically symmetric profiles. Our results suggest that with a different morphology of the matter profile, the spontaneous scalarization may occur more easily. To retain consistency to cosmology it is important to take into account the bare mass of the scalar field. While the analysis in the main text focused on the massless scalar field and a detailed analysis of the massive case is beyond the scope of the present paper, here we provide a simple estimation that our statement is expected to be robust even with the bare mass of the scalar field. From the result of \S\ref{sec:wall}, the critical value for the planar symmetric case is expected to be obtained approximately by the condition $m^2_{\rm eff}=0$, i.e.\ $|\hat\beta_{0,{\rm c}}|= O(m^2\Mpl^2/\rho)$. The allowed range of the mass of the scalar field is $10^{-15} \text{eV} \lesssim m \lesssim 10^{-9} \text{eV}$~\cite{Ramazanoglu:2016kul}. Since the upper bound is determined by guaranteeing the spontaneous scalarization at a spherically symmetric configuration, we do not use it and focus on the lower bound, which translates to
\be
|\hat\beta_{0,{\rm c} }| \gtrsim 
O(1)\mk{ \f{\rho}{\text{MeV}^4} }^{-1}\,.
\ee
As an example, a cosmologically viable model of a domain wall considered in \cite{Olive:2010vh} has energy density $\rho_0\sim O(\text{MeV}^4)$ and the width $d\sim O(\text{MeV}^{-1})$,  for which $|\hat\beta_{0,{\rm c} }|\gtrsim O(1)$. It corresponds to the case with $\hat d\equiv 
d (8\pi G\rho_0/3)^{1/2}
\sim 10^{-21}$. 
On the other hand, for the spherically symmetric configuration, since the critical value for the massless scalar case is $|\hat\beta_{0,{\rm c} }|\sim 10^{10}$ for $\hat d=10^{-5}$, 
the massive scalar case for $\hat d\sim 10^{-21}$ is at least expected to be $|\hat\beta_{0,{\rm c} }| \gg 10^{10}$ 
(see Fig.~\ref{fig:Uath-sph}). 
Further, if we extrapolate Fig.~\ref{fig:Uath-sph} using the scaling relation $|\hat\beta_{0,{\rm c} }|\propto \hat d^{-2}$, it is expected that $|\hat\beta_{0,{\rm c} }|\sim 10^{42}$ for $\hat d=10^{-21}$ for the massless case, and hence the massive case is expected to have $|\hat\beta_{0,{\rm c} }|>10^{42}$.
Therefore, this estimation suggests that in the context of scalar-tensor theories, the shape dependence of the spontaneous scalarization should be taken into account to understand the properties of domain walls.

There are other interesting generalizations of the result of the present paper. In this paper, as the simplest and extreme cases, we focused on the spherically and planar symmetric profiles and analytically showed that $\hbc=0$ for the latter case for any matter profile. It is intriguing to consider other morphology of the matter profile, such as a configuration around a rotating object or a string structure created as a topological defect, and to see how the critical value $\hbc$ varies depending on the morphology. 
Similar analysis would also apply to the shape dependence of other variations of the spontaneous scalarization, e.g.\  with coupling to matter via a disformally transformed metric or coupling to the Maxwell invariant.
It would also be interesting to consider an application of our analysis to 
the dynamical scalarization~\cite{Palenzuela:2013hsa,Barausse:2012da} and
the screening mechanisms such as the chameleon mechanism~\cite{Khoury:2003aq} and symmetron mechanism~\cite{Hinterbichler:2010es} since they are also based on the action~\eqref{action}.
It is natural to expect that these mechanisms
possess a similar shape dependence.  We leave these topics as future works.

\acknowledgments

This work was supported in part by Japan Society for the Promotion of Science (JSPS) Grants-in-Aid for Scientific Research (KAKENHI) No.\ JP17H06359 (H.M., S.M.), No.\ JP18K13565 (H.M.), No.\ JP17H02890 (S.M.), and by World Premier International Research Center Initiative (WPI), MEXT, Japan (S.M.).

\appendix

\section{Static planar symmetric spacetime}
\label{sec:sta}

In this Appendix, we consider the static planar symmetric spacetime 
\be \label{bg3}
ds^2 = - a^2(z)dt^2 + b^2(z)(dx^2+dy^2) + dz^2\,,
\ee
and show that there exists a curvature singularity for the step function density profile.

The Einstein equation is given by
\begin{align} \label{Eeq-ab}
G^t_t=
\f{2b''}{b} + \f{b'^2}{b^2}
&= - 8\pi G\rho\,,\notag\\
G^x_x=G^y_y=
\f{a''}{a} + \f{b''}{b} + \f{a' b'}{a b} 
&= 8\pi G P_x\,,\notag\\
G^z_z=
\f{b'^2}{b^2} + \f{2a' b'}{a b} 
&= 8\pi G P_z\,,
\end{align}
For the step function density profile 
\be
\rho(z)=
\begin{cases}
\rho_0\,, & (|z|\leq d)\,, \\
0\,, & (|z|>d)\,,
\end{cases}
\ee
we can derive the following analytic solution
\begin{align} 
a(\hat z)= b(\hat z)^{-1/2}=
\begin{cases}
 \displaystyle
 \mk{\f{\hat z_c-\hat d}{(\hat z_c-\hat z)\cos (3\hat d/2)} }^{1/3}\,, &  (\hat z>\hat d)\,, \\
\cos^{-1/3} (3\hat z/2)\,, & (-\hat d<\hat z<\hat d)\,,\\
\displaystyle \mk{\f{\hat z_c-\hat d}{(\hat z_c+\hat z)\cos (3\hat d/2)} }^{1/3}\,, &  (\hat z<-\hat d)\,,\\
\end{cases}
\end{align}
where $\hat z_c \equiv \hat d + 2/3 \cot(3 \hat d/2)$. We also note that $|a(\hat z)|\to \infty$ and $b(\hat z)\to 0$ for $\hat z\to \pm \hat z_c$, which is different from the time dependent solution considered in \S\ref{sec:wall} as in that case $a(\hat z)= b(\hat z) \to 0$ for $z\to \pm z_h$. Since the solution satisfies the vacuum Einstein equation, $R_{\mu\nu}=0$ for $|z|>d$. Therefore, the Kretschmann scalar $R_{\mu\nu\rho\sigma}R^{\mu\nu\rho\sigma}=C_{\mu\nu\rho\sigma}C^{\mu\nu\rho\sigma}+2R_{\mu\nu}R^{\mu\nu}-R^2/3$ is determined by the contraction of the Weyl tensor, which is given by
\be
C_{\mu\nu\rho\sigma}C^{\mu\nu\rho\sigma}= \f{64}{27 (\hat z - \hat z_c)^4}\,,
\ee
suggesting the curvature singularity at $\hat z = \hat z_c$.

\bibliography{ref-planar}

\end{document}